# MODELING AND STUDY OF THE CERENKOV EFFECT

I.Angelov[c], E.Duverger[a], L.Makovicka[a], E. Malamova[a], A.Mishev[b], J.Stamenov[b]

[a] RMC/LMIT, Université de Franche-Comté, BP 427, 25211 Montbéliard France
[b] INRNE, BAS, 72 Tsarigradsko chausse, 1784 Sofia Bulgaria
[c] SWU Neofit Rilski Blagoevgrad Bulgaria

**Abstract :**

The studies realized in INRNE (Institute for Nuclear Research and Nuclear Energy) particulary in cosmic rays detection and construction of Muonic Cerenkov Telescope in University of Blagoevgrad [1] shows the need to develop a theoretical model based on observed phenomenon and to refine it for the detection system optimisation. The effect was introduced in EGS4 [2] code system. The first simulations were consecrated to different geometry's of water tank in total reflection. The model was compared with experimental data realised with gamma source $^{60}$Co using the telescope. A simple atmospheric model is introduced in EGS4. The comparison between CORSIKA [3] and EGS4 codes was realised.

## 1.Introduction

A Muonic telescope (fig.1) is developed in INRNE and University of Blagoevgrad Bulgaria [1]. The destination is the registration of secondary muons and measuring the cosmic rays variations. The telescope is based on 18 water Cerenkov detectors split in 2 slabs of 3x3 cells.

The dimensions of the tanks are 50x50x12cm. The penetrating muons creates Cerenkov photons unregistered by photomultipliers (fig.2). The telescope is under absorber so the electrons are rejected.

The modelling of the processes and the development of theoretical model based on observed phenomenon is necessary to calibration and system optimisation. Most of the available codes permitting the simulation of Cerenkov effect are limited to one medium and geometry -Corsika code

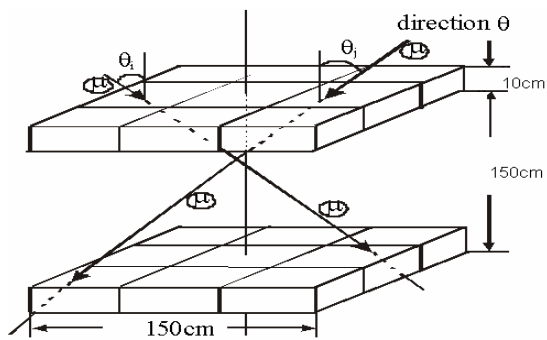  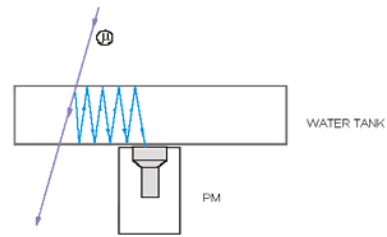

Fig.1 Muonic Cerenkov telescope          Fig.2 Water Cerenkov detector

[3] or simplify the interaction types. On the other hand the experience with EGS4 [2] code system permitting the Monte Carlo simulation of electron-photon showers in different media give us the possibility to use it.

## 2. Theoretical studies

The Cerenkov radiation is emitted if the velocity v of charged particles exceeds the local speed of light which is given by the local refractive index of the medium n and the vacuum speed of light c [4]. So the condition of the effect is

$$nv/c = n\beta > 1 \qquad (1)$$

neglecting the wavelength dependence of n. The angle of emission $\theta_c$ of Cerenkov photons relative to the charged particle direction is

$$\theta_c = \arccos \frac{1}{\beta n} \qquad (2)$$

and the number $N_c$ of photons emitted per path length s in this angle is

$$\frac{dN_c}{ds} = 2\pi\alpha \int \frac{\sin^2 \theta_c}{\lambda^2} d\lambda \qquad (3)$$

In the wavelength band 350-500 nm this gives

$$Ntcher = 390 \sin^2 \theta_c \qquad (4)$$

photons per cm.

The absorption of photons in the medium is not taken into account. The first step was consecrated to simulation of water Cerenkov detector in total reflection. The water is chosen for practical reasons existing detector and the model experimental verification. The photon trajectory is not simulated. The obtained results shows dependence in function of tank geometry and the energy of primary particle corresponding to our expectations because the effect is a treshold phenomenon. The number of charged particles increase slightly with the lateral dimensions of the tank and is from a given energy almost constant (fig.3). So the quantity of Cerenkov photons saturation in function of tank's depth is observed. An additional Monte Carlo code « TRAMEAN » for the mean trajectory calculation and for the detection efficiency optimisation is made. The increasing of photomultiplier surface and decreasing of mean photon path results to efficiency increasing.

**2. Experimental studies**

The experimental studies was realized using a little most efficient tank and gamma source $^{60}$Co. The penetrating in the tank gamma quantum creates Compton electrons which creates Cerenkov photons. For the calibration and experimental verification of the model we need the different quantities of photons produced in the tank. So the response was measured for three depths of water in function of the systems off-set voltage -10 min. per experimental point (fig.4). The research of the working point corresponding to the best efficiency of PM is made using an approximation. On the other hand the total number of Cerenkov photons in the tank is calculated with EGS4. The additional modelling shows 79.2% efficiency of registration. On the other hand the photomultipliers registration efficiency is 10%. Moreover one can see the similar behaviour of experimental and theoretical calculations (fig.4).

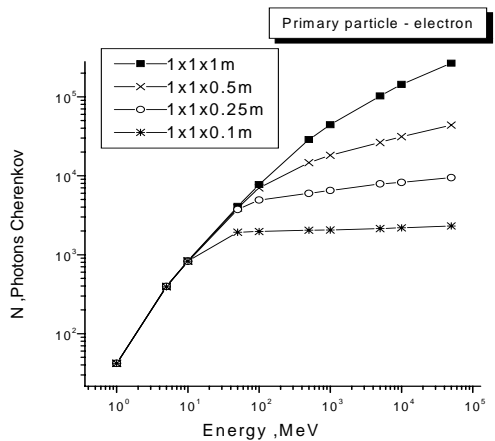

Fig.3 Simulated number of Cherenkov photons in function of primary particle(electron) energy and water tank geometry

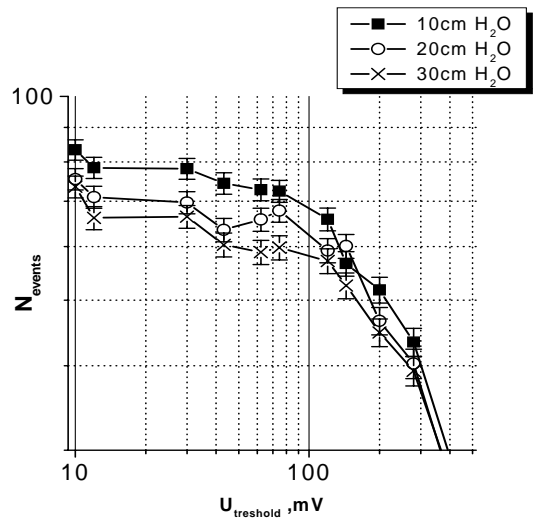

Fig.4 Experimental response of the little tank

The experimental setup is presented in fig. 5. it represents the anticoincidence scheme between the muonic telescope and the water tank. Actually one smaller water tank has been used the aim obtaining three different quantities of Cerenkov photons using different water depths.

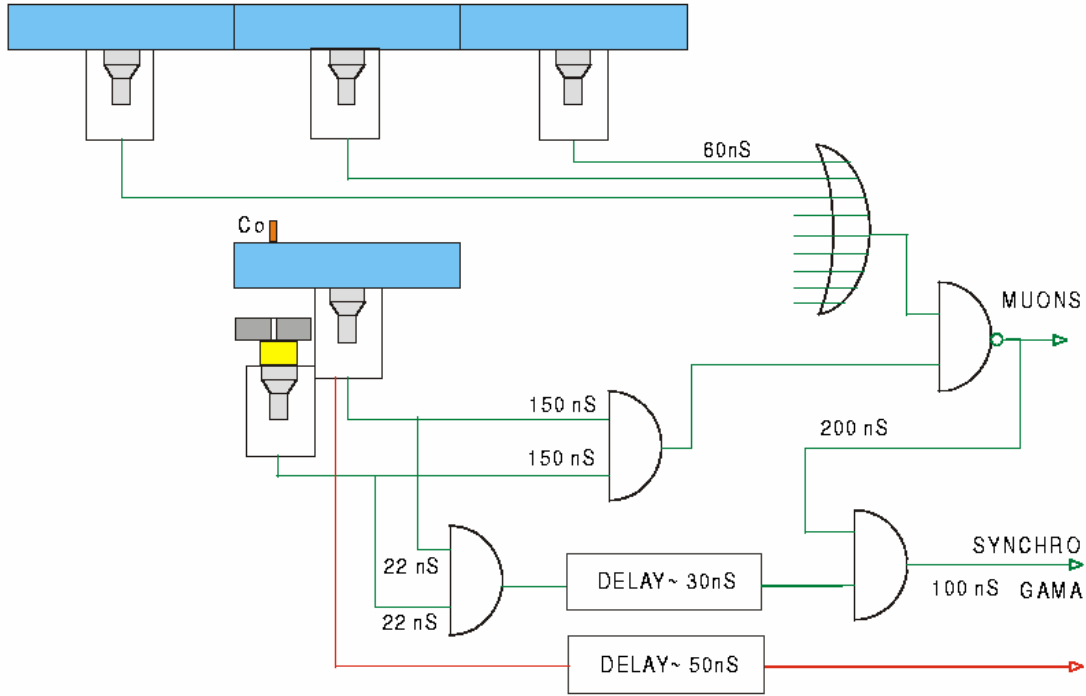

Fig. 5 Experimental setup

## 3. Additional studies

Other possible verification of the model is the comparison with another code for simulation the Cerenkov effect. We have introduced a simple model of the atmosphere in the EGS4. The atmosphere is divided in 21 layers of 5 km thickness. The variation of the refractive index is taken into account. This is important for the lateral distribution function of Cerenkov light. This simple model permits also to track the shower development. The angle of Cerenkov photons emission is also simulated with a full analogy with UPHI subroutine. The comparison with CORSIKA code is realized in large energetic range. We take into account only the total number of Cerenkov photons at sea observation level. The results are shown in (fig.6). So replacing in EGS4 the electron rest mass with a muon rest mass and using our response simulation of the

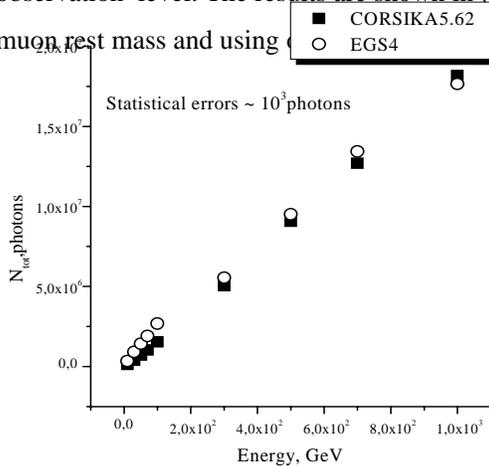

Fig.6 Total number of Cerenkov photons simulated with CORSIKA and EGS4 codes

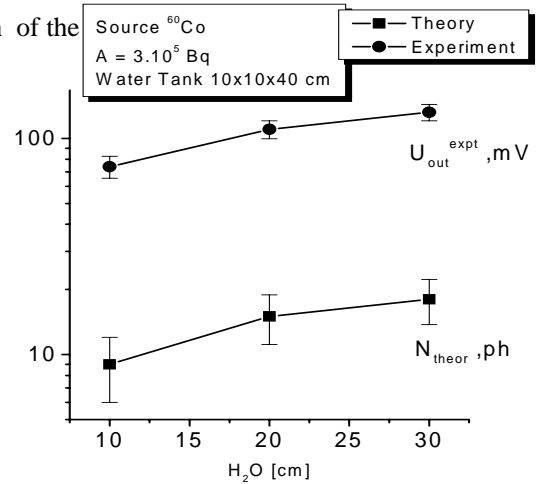

Fig.5 Experimental and theoretical response of the little tank

## 4. Conclusion

The introduction of Cerenkov effect in Egs4 code system was made. This gives a possibility to simulate the response of detectors based in registration of Cerenkov radiation in different media. On the other hand the calibration of existing muon telescope and system registration optimisation is possible. The proposed methodology is based on experimental studies and on Monte Carlo simulation of the response. Our model gives a possibility to simulate the response of detectors based in registration of Cerenkov radiation in different media. On the other hand the calibration of existing muon telescope and system registration optimisation is possible. The proposed methodology is based on experimental studies and on Monte Carlo simulation of the response.